# Tsallis δ-entropy in an accelerating BIon

Hossein Ghaforyan[1,a], Somayeh Shoorvazi[2,b], Alireza Sepehri[2,c], Majid Ebrahimzadeh[3,d]

[1] Department of Photonics, University of Bonab, Bonab 5551761167, Iran
[2] Research Institute for Astronomy and Astrophysics of Maragha (RIAAM), P.O. Box 55134-441, Maragha, Iran
[3] Department of Physics, Payame Noor University (PNU), P.O. Box 19395-3697, Tehran, Iran

**Abstract** In this research, we consider thermodynamically the behaviour of an accelerating BIon and show that the entropy of this system has the form of Tsallis entropy. A BIon is a system that consists of a brane, an anti-brane and a wormhole which connects them together. By increasing the acceleration of branes, the area of BIon increases and its Tsallis entropy grows.

## 1 Introduction

Recently, Tsallis and Cirto have argued that the entropy of a gravitational system such as a black hole could be extended to the non-additive entropy, which is given by $S = \gamma A^\beta$, where A is the horizon area [1]. There have been lots of discussion on this topic so far. For example, some authors have considered the limited behaviour of the evolution of Tsallis entropy in self-gravitating systems. They have shown that the Tsallis entropy generally exhibits a bounded property in self-gravitating system. This indicates the existence of global maximum of Tsallis entropy [2]. Some other authors have proposed a coherence quantifier in terms of the Tsallis relative $\alpha$ entropy which lays the foundation to the non-extensive thermo-statistics and plays the same role as the standard logarithmic entropy does in the information theory [3–5]. In another consideration, authors have derived entropic-force terms from a generalized black-hole entropy proposed by C. Tsallis and L.J.L. Cirto in order to examine the entropic cosmology. Unlike the Bekenstein entropy, which is proportional to area, the generalized entropy is proportional to volume because of appropriate nonadditive generalizations [6]. In another work, the relation between Tsallis entropy and the exchange of energy between the bulk (the universe) and the boundary (the horizon of the universe) has been considered [7]. In another investigation, using Tsallis entropy, the evolution of the universe in entropic cosmologies has been studied. In this model, authors have considered an extended entropic-force model that includes a Hubble parameter (H) term and a constant term in entropic-force terms. The H term is derived from a volume entropy, whereas the constant term is derived from an entropy proportional to the square of an area [8]. In another research, the evolutions of Tsallis entropy during Non-adiabatic-like the accelerated expansion of the late universe has been considered [9]. In addition, the application of this entropy in other aspects of cosmology and physics has been investigated [10,11] in other studies. And finally, employing the modified entropy-area relation suggested by Tsallis and Cirto and the holographic hypothesis, a new holographic dark energy (HDE) model was proposed [12].

In this paper, we will calculate the entropy of an accelerating BIon and will show that this entropy includes some terms similar to Tsallis entropy. A BIon is a configuration which has been constructed from a brane, an anti-brane and a wormhole which connects them together [13–15]. We will discuss that by passing time, the area of BIon increases and the entropy grows.

The outline of the paper is as follows. In Sect. 2, we will obtain the area of a BIon in which branes are boosted with acceleration. In Sect. 3, we will calculate the entropy of an accelerating BIon and show that it is in good agreement with Tsallis entropy. The last section is devoted to conclusion.

## 2 Area of an accelerating BIon with accelerating branes

In this section, we will obtain the area of a BIon with accelerating branes. We will show that the acceleration leads to the emergence of a Rindler space-time. In this space-time, each part of BIon in flat space-time can be transformed to two parts, which act reverse to each other. Each of the parts live in one region (see Figs. 1, 2). When, one part of BIon in region I expands, another part in region II contracts. We will obtain the separation distance between branes in each region and calculate the area of BIon.

[a] e-mail: hghaforyan@yahoo.com
[b] e-mail: shoorvazi@gmail.com
[c] e-mail: alireza.sepehri3@gmail.com
[d] e-mail: pasiran@pasiran.com

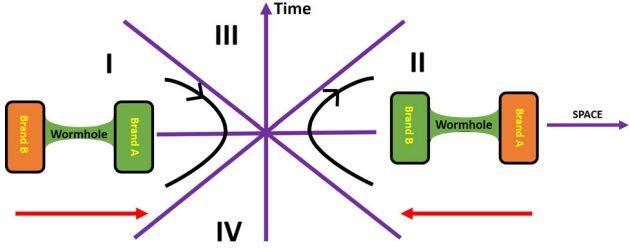

**Fig. 1** Two Flowing BIons with accelerating branes are emerged in two regions I and II of a Rindler horizon

To consider the BIon, we specify an embedding of the D3-brane world volume in 10D Minkowski space-time with metric [18];

$$ds^2 = -dt^2 + dr^2 + r^2\left(d\theta^2 + sin^2\theta d\phi^2\right) + \sum_{i=1}^{6} dx_i^2 \quad (1)$$

without background fluxes. We assume that the branes are boosted with the acceleration. In this case, the relation between the world volume of the coordinates of the accelerating D3-branes ($\tau, \sigma$) and the coordinates of 10D Minkowski space-time ($t, r$) are [19];

$$at = e^{a\sigma}\sinh(a\tau) \quad ar = e^{a\sigma}\cosh(a\tau) \quad \text{In Region I}$$
$$at = -e^{-a\sigma}\sinh(a\tau) \quad ar = e^{-a\sigma}\cosh(a\tau) \quad \text{In Region II} \quad (2)$$

where $a$ is the acceleration of branes. The acceleration leads to the emergence of two new regions in a Rindler space-time. In each region, we have a flowing BIon with two accelerating branes. The behaviour of the BIon in region I is reverse to the BIon in region II. Infact, when one accelerating brane in region I is expanding, its partner in region II is compacting with the same acceleration (see Fig. 1). We can rewrite the Eq. (1) as;

$$ds_I^2 = -dt^2 + dr^2 + r^2\left(d\theta^2 + sin^2\theta d\phi^2\right)$$
$$+ \sum_{i=1}^{6} dx_i^2 = e^{2a\sigma}\left(d\tau^2 - d\sigma^2\right)$$
$$+ \left(\frac{1}{a}e^{a\sigma}\cosh(a\tau)\right)^2\left(d\theta^2 + sin^2\theta d\phi^2\right) + \sum_{i=1}^{6} dx_i^2 \quad (3)$$

$$ds_{II}^2 = -dt^2 + dr^2 + r^2\left(d\theta^2 + sin^2\theta d\phi^2\right)$$
$$+ \sum_{i=1}^{6} dx_i^2 = e^{-2a\sigma}\left(d\tau^2 - d\sigma^2\right)$$
$$+ \left(\frac{1}{a}e^{-a\sigma}\cosh(a\tau)\right)^2\left(d\theta^2 + sin^2\theta d\phi^2\right) + \sum_{i=1}^{6} dx_i^2 \quad (4)$$

We can suppose that the coordinate along the separation distance between branes ($x^4 = z$) depends on the $r = \pm\frac{1}{a}e^{\pm a\sigma}\cosh(a\tau)$ and rewrite the Eqs. (3, 4) as;

$$ds_I^2 = -dt^2 + \left(1 + \left(\frac{dz}{dr}\right)^2\right)dr^2$$
$$+ r^2\left(d\theta^2 + sin^2\theta d\phi^2\right) + \sum_{i=1}^{5} dx_i^2$$
$$= \left(e^{2a\sigma} + \frac{1}{\sinh^2(a\tau)}\left(\frac{dz}{d\tau}\right)^2\right)d\tau^2$$
$$- \left(e^{2a\sigma} + \frac{1}{\cosh^2(a\tau)}\left(\frac{dz}{d\sigma}\right)^2\right)d\sigma^2$$
$$+ \frac{1}{\sinh(a\tau)\cosh(a\tau)}\left(\frac{dz}{d\tau}\frac{dz}{d\sigma}\right)d\tau d\sigma$$
$$+ \left(\frac{1}{a}e^{a\sigma}\cosh(a\tau)\right)^2\left(d\theta^2 + sin^2\theta d\phi^2\right) + \sum_{i=1}^{5} dx_i^2 \quad (5)$$

$$ds_{II}^2 = -dt^2 + \left(1 + \left(\frac{dz}{dr}\right)^2\right)dr^2$$
$$+ r^2\left(d\theta^2 + sin^2\theta d\phi^2\right) + \sum_{i=1}^{5} dx_i^2$$
$$= \left(e^{-2a\sigma} + \frac{1}{\sinh^2(a\tau)}\left(\frac{dz}{d\tau}\right)^2\right)d\tau^2$$
$$- \left(e^{-2a\sigma} + \frac{1}{\cosh^2(a\tau)}\left(\frac{dz}{d\sigma}\right)^2\right)d\sigma^2$$
$$- \frac{1}{\sinh(a\tau)\cosh(a\tau)}\left(\frac{dz}{d\tau}\frac{dz}{d\sigma}\right)d\tau d\sigma$$
$$+ \left(\frac{1}{a}e^{-a\sigma}\cosh(a\tau)\right)^2\left(d\theta^2 + sin^2\theta d\phi^2\right) + \sum_{i=1}^{5} dx_i^2 \quad (6)$$

These equations show that the acceleration of branes leads to the emergence of a new Rindler space-time. In fact, the initial orthogonal metric converts to a non-diagonal metric and one curved space-time is emerged. We will show that this metric produces two types of solutions for BIons. In one region of the Rindler space-time, some energy is transmitted from one brane to another. Consequently, one brane is compacted and another is expanded. While, in another region of the Rindler space-time, the transferring of the energy is reversed and the first parane is expanded and another one is compacted. To show this, we begin with the DBI action for the D3-brane which takes the form [13–15];

$$I_{DBI} = -T_{D3} \int_{w.v.} d^4x \sqrt{-det(\gamma_{ab} + 2\pi l_s^2 F_{ab})}$$
$$+ T_{D3} \int_{w.v.} P[C_{(4)}] \quad (7)$$

where the integrals are performed over the four-dimensional world-volume. Here we have defined the induced world-volume metric;

$$\gamma_{ab} = g_{\mu\nu} \partial_a X^\mu \partial_b X^\nu \quad (8)$$

where $g_{\mu\nu}$ is the background metric, $X^\mu(x^a)$ is the embedding of the brane in the background, $x^a$ is the world-volume coordinates, $a, b = 0, 1, 2, 3$ are world-volume indices and $\mu, \nu = 0, 1, ..., 9$ are target space indices. Furthermore, $F_{ab}$ is the two-form field strength which lives on the brane, $C_{(4)}$ is the RR-four form gauge field of the background, $P[C_{(4)}]$ is its pull-back to the world-volume and $T_{D3}$ is the tension of D3-brane. For flat space-time, above action yields the following Lagrangian;

$$\gamma_{ab} dx^a dx^b = -dt^2 + \left(1 + \left(\frac{dz}{dr}\right)^2\right) dr^2$$
$$+ r^2 \left(d\theta^2 + \sin^2\theta d\phi^2\right)$$
$$\Longrightarrow L = -4\pi T_{D3} \int_{r_0}^\infty dr\, r^2$$
$$\times \sqrt{1 + \left(\frac{dz}{dr}\right)^2 - (2\pi l_s^2 F_{ab})^2} \quad (9)$$

Also, for Rindler space-time, we obtain;

The above equations show that the accelerating boost produces two types of Lagrangians, related to spaces I and II. An observer lives on the branes of a BIon in region I and measures a different Lagrangian in respect to the observer who lives on the branes of a BIon in region II. Thus, the behaviour of fields and also the evolutions of branes in region I are reverse in respect to the evolutions of branes in region II.

On the other hand, the above Lagrangians represent the evolutions of half of the Flowing BIon. This part includes a brane and half of a wormhole. The behaviour of the second part is reverse to this one (see Fig. 2). For example, when one brane compacts, gives its energy to another part and leads to its expansion. Also, we have shown that the behaviour of the partner of each part in the other region of Rindler horizon is reverse to it. Thus, we can obtain the following relations;

$$L_{I,half-A} = L_{II,half-B}$$
$$L_{I,half-B} = L_{II,half-A} \quad (12)$$

Now, we can focus on the evolution of Hamiltonians. Previously, it has been shown that the Hamiltonian of BIon in flat space-time and the metric of Eq. (1) is [13–15];

$$H_{DBI} = 4\pi T_{D3} \int dr \sqrt{1 + \left(\frac{\partial z}{\partial r}\right)^2} F_{DBI}$$
$$F_{DBI} = r^2 \sqrt{1 + \frac{K^2}{r^4}} \quad (13)$$

For the Rindler space-time and coordinates in Eq. (2) and also Lagrangians in Eqs. (10, 11), we can obtain the below Hamiltonians;

$$\gamma_{ab,I} dx^a dx^b = \left(e^{2a\sigma} + \frac{1}{\sinh^2(a\tau)}\left(\frac{dz}{d\tau}\right)^2\right) d\tau^2 - \left(e^{2a\sigma} + \frac{1}{\cosh^2(a\tau)}\left(\frac{dz}{d\sigma}\right)^2\right) d\sigma^2$$
$$+ \frac{1}{\sinh(a\tau)\cosh(a\tau)}\left(\left(\frac{dz}{d\tau}\frac{dz}{d\sigma}\right)\right) d\tau d\sigma + \left(\frac{1}{a}e^{a\sigma}\cosh(a\tau)\right)^2 (d\theta^2 + \sin^2\theta d\phi^2) \Longrightarrow L_{I,half}$$
$$= -T_{D3} \int_{\sigma_0}^\infty d\sigma \left(\frac{1}{a}e^{a\sigma}\cosh(a\tau)\right)^2 (\sinh^2(a\tau) + \cosh^2(a\tau))$$
$$\times \sqrt{1 + \frac{e^{-2a\sigma}}{\sinh^2(a\tau)}\left(\frac{dz}{d\tau}\right)^2 + \frac{e^{-2a\sigma}}{\cosh^2(a\tau)}\left(\frac{dz}{d\sigma}\right)^2 + \frac{e^{-2a\sigma}}{\sinh(a\tau)\cosh(a\tau)}\left(\left(\frac{dz}{d\tau}\frac{dz}{d\sigma}\right)\right) - (2\pi l_s^2 F_{ab})^2} \quad (10)$$

$$\gamma_{ab,II} dx^a dx^b = \left(e^{-2a\sigma} + \frac{1}{\sinh^2(a\tau)}\left(\frac{dz}{d\tau}\right)^2\right) d\tau^2 - \left(e^{-2a\sigma} + \frac{1}{\cosh^2(a\tau)}\left(\frac{dz}{d\sigma}\right)^2\right) d\sigma^2$$
$$- \frac{1}{\sinh(a\tau)\cosh(a\tau)}\left(\frac{dz}{d\tau}\frac{dz}{d\sigma}\right) d\tau d\sigma + \left(\frac{1}{a}e^{-a\sigma}\cosh(a\tau)\right)^2 (d\theta^2 + \sin^2\theta d\phi^2) \Longrightarrow L_{II,half}$$
$$= -T_{D3} \int_{\sigma_0}^\infty d\sigma \left(\frac{1}{a}e^{-a\sigma}\cosh(a\tau)\right)^2 (\sinh^2(a\tau) + \cosh^2(a\tau))$$
$$\times \sqrt{1 + \frac{e^{2a\sigma}}{\sinh^2(a\tau)}\left(\frac{dz}{d\tau}\right)^2 + \frac{e^{2a\sigma}}{\cosh^2(a\tau)}\left(\frac{dz}{d\sigma}\right)^2 - \frac{e^{2a\sigma}}{\sinh(a\tau)\cosh(a\tau)}\left(\frac{dz}{d\tau}\frac{dz}{d\sigma}\right) - (2\pi l_s^2 F_{ab})^2} \quad (11)$$

**Fig. 2** Two parts of the Flowing BIon acts reverse to each other and similar to partner of another part in other region. Parts with the same color behaves similar to each other

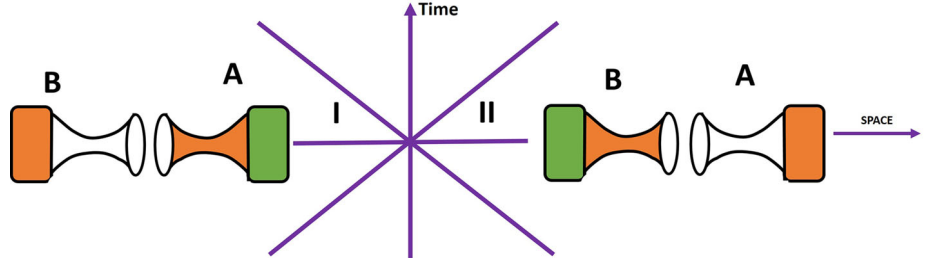

$$H_{DBI,I,half} = T_{D3} \int d\sigma \sqrt{1 + \frac{e^{-2a\sigma}}{\sinh^2(a\tau)}\left(\frac{dz}{d\tau}\right)^2 + \frac{e^{-2a\sigma}}{\cosh^2(a\tau)}\left(\frac{dz}{d\sigma}\right)^2 + \frac{e^{-2a\sigma}}{\sinh(a\tau)\cosh(a\tau)}\left(\left(\frac{dz}{d\tau}\frac{dz}{d\sigma}\right)\right)} F_{DBI,II,half}$$

$$F_{DBI,I,half} = \left(\frac{1}{a}e^{a\sigma}\cosh(a\tau)\right)^2 \sqrt{1 + \frac{K^2}{\left(\frac{1}{a}e^{a\sigma}\cosh(a\tau)\right)^4}\left(\sinh^2(a\tau) + cosh^2(a\tau)\right)} \quad (14)$$

$$H_{DBI,II,half} = T_{D3} \int d\sigma \sqrt{1 + \frac{e^{2a\sigma}}{\sinh^2(a\tau)}\left(\frac{dz}{d\tau}\right)^2 + \frac{e^{2a\sigma}}{\cosh^2(a\tau)}\left(\frac{dz}{d\sigma}\right)^2 - \frac{e^{2a\sigma}}{\sinh(a\tau)\cosh(a\tau)}\left(\left(\frac{dz}{d\tau}\frac{dz}{d\sigma}\right)\right)} F_{DBI,II,half}$$

$$F_{DBI,II,half} = \left(\frac{1}{a}e^{-a\sigma}\cosh(a\tau)\right)^2 \sqrt{1 + \frac{K^2}{\left(\frac{1}{a}e^{-a\sigma}\cosh(a\tau)\right)^4}\left(\sinh^2(a\tau) + cosh^2(a\tau)\right)} \quad (15)$$

These equations describe the evolutions of the energy of the Flowing BIons in region I and region II. It is clear that both Hamiltonians depend on the acceleration, time and coordinate of branes. Also, when the Hamiltonian of one system grows, the Hamiltonian of the other system in another region decreases.

Figure 2 shows that each Flowing BIon is divided into two parts in each region. These parts act reverse to each other. For example, when one part expands, the other part compacts. This is because the energy is transmitted from one brane to another and consequently, the first brane compacts and another one expands. On the other hand, each part acts reverse to the partner in another region. Thus, we can write the below relations between Hamiltonians;

$$H_{I,half-A} = H_{II,half-B}$$
$$H_{I,half-B} = H_{II,half-A} \quad (16)$$

The wave equations which are extracted from equations (14, 15) are;

$$\frac{\partial}{\partial \tau}\left(\frac{\frac{e^{-2a\sigma}}{\sinh^2(a\tau)}\left(\frac{dz}{d\tau}\right)F_{DBI,I,A}}{\sqrt{1 + \frac{e^{-2a\sigma}}{\sinh^2(a\tau)}\left(\frac{dz}{d\tau}\right)^2 + \frac{e^{-2a\sigma}}{\cosh^2(a\tau)}\left(\frac{dz}{d\sigma}\right)^2 + \frac{e^{-2a\sigma}}{\sinh(a\tau)\cosh(a\tau)}\left(\left(\frac{dz}{d\tau}\frac{dz}{d\sigma}\right)\right)}}\right)$$
$$+ \frac{\partial}{\partial \sigma}\left(\frac{\frac{e^{-2a\sigma}}{\cosh^2(a\tau)}\left(\frac{dz}{d\sigma}\right)F_{DBI,I,A}}{\sqrt{1 + \frac{e^{-2a\sigma}}{\sinh^2(a\tau)}\left(\frac{dz}{d\tau}\right)^2 + \frac{e^{-2a\sigma}}{\cosh^2(a\tau)}\left(\frac{dz}{d\sigma}\right)^2 + \frac{e^{-2a\sigma}}{\sinh(a\tau)\cosh(a\tau)}\left(\left(\frac{dz}{d\tau}\frac{dz}{d\sigma}\right)\right)}}\right)$$
$$+ \frac{\partial^2}{\partial \sigma \partial \tau}\left(\frac{\frac{e^{-2a\sigma}}{\sinh(a\tau)\cosh(a\tau)}\left(\frac{dz}{d\tau}+\frac{dz}{d\sigma}\right)F_{DBI,I,A}}{\sqrt{1 + \frac{e^{-2a\sigma}}{\sinh^2(a\tau)}\left(\frac{dz}{d\tau}\right)^2 + \frac{e^{-2a\sigma}}{\cosh^2(a\tau)}\left(\frac{dz}{d\sigma}\right)^2 + \frac{e^{-2a\sigma}}{\sinh(a\tau)\cosh(a\tau)}\left(\left(\frac{dz}{d\tau}\frac{dz}{d\sigma}\right)\right)}}\right) = 0 \quad (17)$$

$$\frac{\partial}{\partial \tau}\left(\frac{\frac{e^{2a\sigma}}{\sinh^2(a\tau)}\left(\frac{dz}{d\tau}\right)F_{DBI,I,B}}{\sqrt{1+\frac{e^{2a\sigma}}{\sinh^2(a\tau)}\left(\frac{dz}{d\tau}\right)^2+\frac{e^{2a\sigma}}{\cosh^2(a\tau)}\left(\frac{dz}{d\sigma}\right)^2-\frac{e^{2a\sigma}}{\sinh(a\tau)\cosh(a\tau)}\left(\left(\frac{dz}{d\tau}\frac{dz}{d\sigma}\right)\right)}}\right)$$
$$+\frac{\partial}{\partial \sigma}\left(\frac{\frac{e^{2a\sigma}}{\cosh^2(a\tau)}\left(\frac{dz}{d\sigma}\right)F_{DBI,I,B}}{\sqrt{1+\frac{e^{2a\sigma}}{\sinh^2(a\tau)}\left(\frac{dz}{d\tau}\right)^2+\frac{e^{2a\sigma}}{\cosh^2(a\tau)}\left(\frac{dz}{d\sigma}\right)^2-\frac{e^{2a\sigma}}{\sinh(a\tau)\cosh(a\tau)}\left(\left(\frac{dz}{d\tau}\frac{dz}{d\sigma}\right)\right)}}\right)$$
$$+\frac{\partial^2}{\partial\sigma\partial\tau}\left(\frac{\frac{e^{2a\sigma}}{\sinh(a\tau)\cosh(a\tau)}\left(\frac{dz}{d\tau}+\frac{dz}{d\sigma}\right)F_{DBI,I,B}}{\sqrt{1+\frac{e^{2a\sigma}}{\sinh^2(a\tau)}\left(\frac{dz}{d\tau}\right)^2+\frac{e^{2a\sigma}}{\cosh^2(a\tau)}\left(\frac{dz}{d\sigma}\right)^2-\frac{e^{2a\sigma}}{\sinh(a\tau)\cosh(a\tau)}\left(\left(\frac{dz}{d\tau}\frac{dz}{d\sigma}\right)\right)}}\right)=0 \quad (18)$$

By solving the above equations, we can obtain the below form for the separation distance between branes;

$$z_{I-A}=z_{II-B}\simeq\int d\tau d\sigma\left(e^{-4a\sigma}\sinh^2(a\tau)\cosh^2(a\tau)\right)$$
$$\times\left(\frac{F_{DBI,I,A}(\tau,\sigma)\left(\frac{F_{DBI,I,A}(\tau,\sigma)}{F_{DBI,I,A}(\tau,\sigma_0)}-e^{-4a(\sigma-\sigma_0)}\frac{\cosh^2(a\tau_0)}{\cosh^2(a\tau)}\right)^{-\frac{1}{2}}}{F_{DBI,I,A}(\tau_0,\sigma)\left(\frac{F_{DBI,I,A}(\tau_0,\sigma)}{F_{DBI,I,A}(\tau_0,\sigma_0)}-e^{-4a(\sigma-\sigma_0)}\frac{\cosh^2(a\tau_0)}{\cosh^2(a\tau)}\right)^{-\frac{1}{2}}}-\frac{\sinh^2(a\tau_0)}{\sinh^2(a\tau)}\right)^{-\frac{1}{2}} \quad (19)$$

$$z_{I-B}=z_{II-A}\simeq\int d\tau d\sigma\left(e^{4a\sigma}\sinh^2(a\tau)\cosh^2(a\tau)\right)$$
$$\times\left(\frac{F_{DBI,II,A}(\tau,\sigma)\left(\frac{F_{DBI,II,A}(\tau,\sigma)}{F_{DBI,II,A}(\tau,\sigma_0)}-e^{4a(\sigma-\sigma_0)}\frac{\cosh^2(a\tau_0)}{\cosh^2(a\tau)}\right)^{-\frac{1}{2}}}{F_{DBI,II,A}(\tau_0,\sigma)\left(\frac{F_{DBI,II,A}(\tau_0,\sigma)}{F_{DBI,II,A}(\tau_0,\sigma_0)}-e^{4a(\sigma-\sigma_0)}\frac{\cosh^2(a\tau_0)}{\cosh^2(a\tau)}\right)^{-\frac{1}{2}}}-\frac{\sinh^2(a\tau_0)}{\sinh^2(a\tau)}\right)^{-\frac{1}{2}} \quad (20)$$

The total separation between branes in the two regions are;

$$z_I = z_{I-A} + z_{I-B}$$
$$z_{II} = z_{II-A} + z_{II-B} \implies$$
$$z_I = z_{II} \quad (21)$$

The above equation shows that the total separation distance between branes depends on the acceleration, time and coordinates of branes and is independent of regions. This is because when the separation distance for a part of BIon grows, this parameter in another part decreases. Also, the separation of distance for each part in another region acts reversely and thus, the total separation distance in both regions acts similarly.

Now, we can calculate the area of BIon. In our consideration, BIon has a cylindrical shape and thus, we can obtain:

$$A=\pi r^2 z_I=\pi\left(\frac{1}{a}e^{\pm a\sigma}\cosh(a\tau)\right)^2\left[\left(\int d\tau d\sigma\left(e^{-4a\sigma}\sinh^2(a\tau)\cosh^2(a\tau)\right)\right.\right.$$
$$\times\left(\frac{F_{DBI,I,A}(\tau,\sigma)\left(\frac{F_{DBI,I,A}(\tau,\sigma)}{F_{DBI,I,A}(\tau,\sigma_0)}-e^{-4a(\sigma-\sigma_0)}\frac{\cosh^2(a\tau_0)}{\cosh^2(a\tau)}\right)^{-\frac{1}{2}}}{F_{DBI,I,A}(\tau_0,\sigma)\left(\frac{F_{DBI,I,A}(\tau_0,\sigma)}{F_{DBI,I,A}(\tau_0,\sigma_0)}-e^{-4a(\sigma-\sigma_0)}\frac{\cosh^2(a\tau_0)}{\cosh^2(a\tau)}\right)^{-\frac{1}{2}}}-\frac{\sinh^2(a\tau_0)}{\sinh^2(a\tau)}\right)^{-\frac{1}{2}}\right)$$
$$+\left(\int d\tau d\sigma\left(e^{4a\sigma}\sinh^2(a\tau)\cosh^2(a\tau)\right)\right.$$
$$\left.\left.\times\left(\frac{F_{DBI,II,A}(\tau,\sigma)\left(\frac{F_{DBI,II,A}(\tau,\sigma)}{F_{DBI,II,A}(\tau,\sigma_0)}-e^{4a(\sigma-\sigma_0)}\frac{\cosh^2(a\tau_0)}{\cosh^2(a\tau)}\right)^{-\frac{1}{2}}}{F_{DBI,II,A}(\tau_0,\sigma)\left(\frac{F_{DBI,II,A}(\tau_0,\sigma)}{F_{DBI,II,A}(\tau_0,\sigma_0)}-e^{4a(\sigma-\sigma_0)}\frac{\cosh^2(a\tau_0)}{\cosh^2(a\tau)}\right)^{-\frac{1}{2}}}-\frac{\sinh^2(a\tau_0)}{\sinh^2(a\tau)}\right)^{-\frac{1}{2}}\right)\right] \quad (22)$$

Above equation shows total area of BIon. This area depends on the acceleration of BIon and time. By passing time and increasing the acceleration of BIon, the separation distance between branes increases and the area grows. On the other hand, for negative accelerations, the separation distance between branes decreases and the area decreases.

## 3 Tsallis entropy of an accelerating BIon

In this section, we will obtain the entropy of accelerating BIon and compare it with Tsallis entropy. We will show that the accelerating BIon produces exactly Tsallis entropy. To this aim, we can turn to obtaining the thermodynamics for a Flowing BIon with accelerating branes, at non-zero temperature. We can obtain this from a black D3-F1 brane bound state geometry that is placed in a Rindler space-time. We begin with the D3-F1 black brane bound state background in flat space-time which has the string frame metric [20]

$$ds^2 = D^{-\frac{1}{2}} H^{-\frac{1}{2}} \left(dx_2^2 + dx_3^2\right) + D^{\frac{1}{2}} H^{-\frac{1}{2}} \left(-fdt^2 + dx_1^2\right) + D^{-\frac{1}{2}} H^{\frac{1}{2}} \left(f^{-1} dr^2 + r^2 d\Omega_5^2\right) \quad (23)$$

where

$$f = 1 - \frac{r_0^4}{r^4} \quad H = 1 + \frac{r_0^4 \sinh^2 \alpha}{r^4}$$
$$D = \cos^2 \epsilon + \sin^2 \epsilon H^{-1} \quad (24)$$

And

$$\cosh^2 \alpha = \frac{3}{2} \frac{\cos \frac{\delta}{3} + \sqrt{3} \cos \frac{\delta}{3}}{\cos \delta}$$
$$\cos \epsilon = \frac{1}{\sqrt{1 + \frac{K^2}{r^4}}} \quad (25)$$

With the definition;

$$\cos \delta = \bar{T}^4 \sqrt{1 + \frac{K^2}{r^4}}$$
$$\bar{T} = \left(\frac{9\pi^2 N}{4\sqrt{3} T_{D3}}\right)^{\frac{1}{2}} T \quad (26)$$

Comparing the metric in Eq. (23) with the metrics of Eqs. (5, 6), we obtain the explicit form of metrics of thermal BIons in region I and region II of the Rindler space-time;

$$ds^2_{I,A,thermal}$$
$$= D^{\frac{1}{2}}_{I-A} H^{-\frac{1}{2}}_{I-A} f_{I-A} \left(e^{2a\sigma} + \frac{1}{\sinh^2(a\tau)} \left(\frac{dz}{d\tau}\right)^2\right) d\tau^2$$
$$- D^{-\frac{1}{2}}_{I-A} H^{\frac{1}{2}}_{I-A} f^{-1}_{I-A} \left(e^{2a\sigma} + \frac{1}{\cosh^2(a\tau)} \left(\frac{dz}{d\sigma}\right)^2\right) d\sigma^2$$
$$+ \frac{1}{\sinh(a\tau)\cosh(a\tau)} \left(\frac{dz}{d\tau} \frac{dz}{d\sigma}\right) d\tau d\sigma$$
$$+ D^{-\frac{1}{2}}_{I-A} H^{\frac{1}{2}}_{I-A} \left(\frac{1}{a} e^{a\sigma} \cosh(a\tau)\right)^2 \left(d\theta^2 + \sin^2\theta d\phi^2\right)$$
$$+ D^{-\frac{1}{2}}_{I-A} H^{-\frac{1}{2}}_{I-A} \sum_{i=1}^{5} dx_i^2 \quad (27)$$

$$ds^2_{II,A,thermal}$$
$$= D^{\frac{1}{2}}_{II-A} H^{-\frac{1}{2}}_{II-A} f_{II-A} \left(e^{-2a\sigma} + \frac{1}{\sinh^2(a\tau)} \left(\frac{dz}{d\tau}\right)^2\right) d\tau^2$$
$$- D^{-\frac{1}{2}}_{II-A} H^{\frac{1}{2}}_{II-A} f^{-1}_{II-A} \left(e^{-2a\sigma} + \frac{1}{\cosh^2(a\tau)} \left(\frac{dz}{d\sigma}\right)^2\right) d\sigma^2$$
$$- \frac{1}{\sinh(a\tau)\cosh(a\tau)} (\frac{dz}{d\tau} \frac{dz}{d\sigma}) d\tau d\sigma$$
$$+ D^{-\frac{1}{2}}_{II-A} H^{\frac{1}{2}}_{II-A} \left(\frac{1}{a} e^{-a\sigma} \cosh(a\tau)\right)^2 \left(d\theta^2 + \sin^2\theta d\phi^2\right)$$
$$+ D^{-\frac{1}{2}}_{II-A} H^{-\frac{1}{2}}_{II-A} \sum_{i=1}^{5} dx_i^2 \quad (28)$$

where

$$f_{I-A} = 1 - \frac{(e^{a\sigma_0} \cosh(a\tau_0))^4}{(e^{a\sigma} \cosh(a\tau))^4} \quad H_{I-A} = 1$$
$$+ \frac{(e^{a\sigma_0} \cosh(a\tau_0))^4 \sinh^2 \alpha_{I-A}}{(e^{a\sigma} \cosh(a\tau))^4}$$
$$D_{I-A} = \cos^2 \epsilon_{I-A} + \sin^2 \epsilon_{I-A} H^{-1}_{I-A} \quad (29)$$
$$f_{II-A} = 1 - \frac{\left(e^{-a\sigma_0} \cosh(a\tau_0)\right)^4}{\left(e^{-a\sigma} \cosh(a\tau)\right)^4} \quad H_{II-A} = 1$$
$$+ \frac{(e^{a\sigma_0} \cosh(a\tau_0))^4 \sinh^2 \alpha_{II-A}}{(e^{a\sigma} \cosh(a\tau))^4}$$
$$D_{II-A} = \cos^2 \epsilon_{II-A} + \sin^2 \epsilon_{II-A} H^{-1}_{II-A} \quad (30)$$

And

$$\cosh^2 \alpha_{I-A} = \frac{3}{2} \frac{\cos \frac{\delta_{I-A}}{3} + \sqrt{3} \cos \frac{\delta_{I-A}}{3}}{\cos \delta_{I-A}}$$
$$\cos \epsilon_{I-A} = \frac{1}{\sqrt{1 + \frac{K^2}{(a^{-1} e^{-a\sigma} \cosh(a\tau))^4}}} \quad (31)$$
$$\cosh^2 \alpha_{II-A} = \frac{3}{2} \frac{\cos \frac{\delta_{II-A}}{3} + \sqrt{3} \cos \frac{\delta_{II-A}}{3}}{\cos \delta_{II-A}}$$
$$\cos \epsilon_{II-A} = \frac{1}{\sqrt{1 + \frac{K^2}{(a^{-1} e^{a\sigma} \cosh(a\tau))^4}}} \quad (32)$$

With the definition;

$$\cos \delta_{I-A} = \bar{T}^4_{0,I-A} \sqrt{1 + \frac{K^2}{(a^{-1} e^{-a\sigma} \cosh(a\tau))^4}}$$

$$\bar{T}_{0,I-A} = \left(\frac{9\pi^2 N}{4\sqrt{3}T_{D3}}\right)^{\frac{1}{2}} T_{0,I-A} \quad (33)$$

$$\cos\delta_{II-A} = \bar{T}_{0,II-A}^4 \sqrt{1 + \frac{K^2}{(a^{-1}e^{a\sigma}\cosh(a\tau))^4}}$$

$$\bar{T}_{0,II-A} = \left(\frac{9\pi^2 N}{4\sqrt{3}T_{D3}}\right)^{\frac{1}{2}} T_{0,II-A} \quad (34)$$

where $T_0$ is the temperature of BIon in non-Rindler space-time. The above results show that how the metric of the D3-F1 black brane evolves by the acceleration and time in a Rindler space-time. It is observed that the thermal BIon which is constructed by black D3-branes experiences different phase transitions by growing or decreasing the acceleration and temperature. It is clear that each BIon is divided into two parts, one expands and the other contracts. Reversely, their partners in region II contract and expand (see Fig. 2).

Previously, thermodynamics of a thermal BIon in flat space-time has been considered and leads to the following results;

$$M = \frac{4T_{D3}^2}{\pi T^4} \int_{r_0}^{\infty} dr \frac{F(r)r^2}{\sqrt{F^2(r) - F^2(r_0)}} \frac{4\cosh^2\alpha + 1}{\cosh^4\alpha}$$

$$S = \frac{4T_{D3}^2}{\pi T^5} \int_{r_0}^{\infty} dr \frac{F(r)r^2}{\sqrt{F^2(r) - F^2(r_0)}} \frac{4}{\cosh^4\alpha}$$

$$Free = \frac{4T_{D3}^2}{\pi T^4} \int_{r_0}^{\infty} dr \sqrt{1 + (\frac{\partial z}{\partial r})^2} F(r) \quad (36)$$

with new definition of $F(r)$ [13];

$$F(r) = r^2 \frac{4\cosh^2\alpha - 3}{\cosh^4\alpha} \quad (37)$$

where $M$ is the mass, $S$ is the entropy and *Free* is the free energy of system.

For the Rindler space-time with the metrics of Eqs. (27, 28) and coordinates of (2), we obtain;

$$M_{I-A} = M_{II-B} = \frac{4T_{D3}^2}{\pi T_{0,I-A}^4} \int_{\sigma_0}^{\infty} d\sigma \frac{F_{DBI,I,A}(\sigma,\tau)\left(\frac{1}{a}e^{a\sigma}\cosh(a\tau)\right)^2 (\sinh^2(a\tau) + \cosh^2(a\tau))}{\sqrt{F_{DBI,I,A}^2(\sigma,\tau) - F_{DBI,I,A}^2(\sigma_o,\tau)}} \times \frac{4\cosh^2\alpha_{I-A} + 1}{\cosh^4\alpha_{I-A}}$$

$$S_{I-A} = S_{II-B} = \frac{4T_{D3}^2}{\pi T_{0,I-A}^5} \int_{\sigma_0}^{\infty} d\sigma \frac{F_{DBI,I,A}(\sigma,\tau)\left(\frac{1}{a}e^{a\sigma}\cosh(a\tau)\right)^2 (\sinh^2(a\tau) + \cosh^2(a\tau))}{\sqrt{F_{DBI,I,A}^2(\sigma,\tau) - F_{DBI,I,A}^2(\sigma_o,\tau)}} \times \frac{4}{\cosh^4\alpha_{I-A}}$$

$$Free_{I-A} = Free_{II-B} = \frac{4T_{D3}^2}{\pi T_{0,I-A}^4} \int_{r_0}^{\infty} d\sigma F_{DBI,I,A}(\sigma,\tau) \left(\frac{1}{a}e^{a\sigma}\cosh(a\tau)\right)^2 \left(\sinh^2(a\tau) + \cosh^2(a\tau)\right)$$

$$\times \sqrt{1 + \frac{e^{-2a\sigma}}{\sinh^2(a\tau)}\left(\frac{dz}{d\tau}\right)^2 + \frac{e^{-2a\sigma}}{\cosh^2(a\tau)}\left(\frac{dz}{d\sigma}\right)^2 + \frac{e^{-2a\sigma}}{\sinh(a\tau)\cosh(a\tau)}\left(\left(\frac{dz}{d\tau}\frac{dz}{d\sigma}\right)\right)} \quad (38)$$

$$ds_{I,A,thermal}^2 = ds_{II,B,thermal}^2$$
$$ds_{I,B,thermal}^2 = ds_{II,A,thermal}^2 \quad (35)$$

At this stage, we can calculate the total mass, entropy and free energy of a thermal BIon in the Rindler space-time.

with the below definition of $F_{DBI,I,A}$;

$$F_{DBI,I,A} = F_{DBI,II,B}$$
$$= \left(a^{-1}e^{a\sigma}\cosh(a\tau)\right)^2 \frac{4\cosh^2\alpha_{I-A} - 3}{\cosh^4\alpha_{I-A}} \quad (39)$$

And

$$M_{II-A} = M_{I-B} = \frac{4T_{D3}^2}{\pi T_{0,II-A}^4} \int_{\sigma_0}^{\infty} d\sigma \frac{F_{DBI,II,A}(\sigma,\tau)\left(\frac{1}{a}e^{-a\sigma}\cosh(a\tau)\right)^2 (\sinh^2(a\tau) + \cosh^2(a\tau))}{\sqrt{F_{DBI,II,A}^2(\sigma,\tau) - F_{DBI,II,A}^2(\sigma_o,\tau)}} \frac{4\cosh^2\alpha_{II-A} + 1}{\cosh^4\alpha_{II-A}}$$

$$S_{II-A} = S_{I-B} = \frac{4T_{D3}^2}{\pi T_{0,II-A}^5} \int_{\sigma_0}^{\infty} d\sigma \frac{F_{DBI,II,A}(\sigma,\tau)\left(\frac{1}{a}e^{-a\sigma}\cosh(a\tau)\right)^2 (\sinh^2(a\tau) + \cosh^2(a\tau))}{\sqrt{F_{DBI,II,A}^2(\sigma,\tau) - F_{DBI,II,A}^2(\sigma_o,\tau)}} \frac{4}{\cosh^4\alpha_{II-A}}$$

$$Free_{II-A} = Free_{I-B} = \frac{4T_{D3}^2}{\pi T_{0,II-A}^4} \int_{\sigma_0}^{\infty} d\sigma F_{DBI,II,A}(\sigma,\tau) \left(\frac{1}{a}e^{-a\sigma}\cosh(a\tau)\right)^2 \left(\sinh^2(a\tau) + \cosh^2(a\tau)\right)$$

$$\times \sqrt{1 + \frac{e^{2a\sigma}}{\sinh^2(a\tau)}\left(\frac{dz}{d\tau}\right)^2 + \frac{e^{2a\sigma}}{\cosh^2(a\tau)}\left(\frac{dz}{d\sigma}\right)^2 + \frac{e^{2a\sigma}}{\sinh(a\tau)\cosh(a\tau)}\left(\left(\frac{dz}{d\tau}\frac{dz}{d\sigma}\right)\right)} \quad (40)$$

with the below definition of $F_{DBI,II,A}$;

$$F_{DBI,II,A} = F_{DBI,I,B}$$
$$= \left(a^{-1}e^{-a\sigma}\cosh(a\tau)\right)^2 \frac{4\cosh^2\alpha_{II-A} - 3}{\cosh^4\alpha_{II-A}} \quad (41)$$

Above equations describe the thermodynamics of the flowing BIon which its branes are boosted with accelerations. It is clear that when the entropy, mass or free energy of brane grows, the ones of other brane decreases.

Now, we can calculate the total entropy of BIon:

$$S_I = S_{I-A} + S_{I-B} = \frac{4T_{D3}^2}{\pi T_{0,I-A}^5}\int_{\sigma_0}^{\infty} d\sigma \frac{F_{DBI,I,A}(\sigma,\tau)\left(\frac{1}{a}e^{a\sigma}\cosh(a\tau)\right)^2 \left(\sinh^2(a\tau) + \cosh^2(a\tau)\right)}{\sqrt{F_{DBI,I,A}^2(\sigma,\tau) - F_{DBI,I,A}^2(\sigma_o,\tau)}} \frac{4}{\cosh^4\alpha_{I-A}}$$
$$+ \frac{4T_{D3}^2}{\pi T_{0,II-A}^5}\int_{\sigma_0}^{\infty} d\sigma \frac{F_{DBI,II,A}(\sigma,\tau)\left(\frac{1}{a}e^{-a\sigma}\cosh(a\tau)\right)^2 \left(\sinh^2(a\tau) + \cosh^2(a\tau)\right)}{\sqrt{F_{DBI,II,A}^2(\sigma,\tau) - F_{DBI,II,A}^2(\sigma_o,\tau)}} \frac{4}{\cosh^4\alpha_{II-A}} \quad (42)$$

To obtain the explicit form of entropy, we should solve Eqs. (39 and 41) and obtain:

$$\cosh^2\alpha_{I-A}$$
$$= \frac{\left(a^{-1}e^{a\sigma}\cosh(a\tau)\right)^2 + \sqrt{\left(a^{-1}e^{a\sigma}\cosh(a\tau)\right)^4 + 12F_{DBI,I,A}}}{4F_{DBI,I,A}}$$

$$\cosh^2\alpha_{II-A}$$
$$= \frac{\left(a^{-1}e^{-a\sigma}\cosh(a\tau)\right)^2 + \sqrt{\left(a^{-1}e^{-a\sigma}\cosh(a\tau)\right)^4 + 12F_{DBI,II,A}}}{4F_{DBI,II,A}}$$
$$\quad (43)$$

Substituting Eq. (43) in Eq. (42) and comparing with Eq. (22), we can obtain:

$$S_I = S_{II} = \frac{4T_{D3}^2}{\pi T_{0,I-A}^5}\left[1 + \frac{T_{0,I-A}^5}{\pi T_{0,II-A}^5}\right] A^{\frac{3}{2}} \quad (44)$$

This entropy is very similar to Tsallis entropy ($S = \gamma A^\beta$) with ($\gamma = \frac{4T_{D3}^2}{\pi T_{0,I-A}^5}[1 + \frac{T_{0,I-A}^5}{\pi T_{0,II-A}^5}]$ and $\beta = \frac{3}{2}$). This means that Tsallis entropy not only works in black hole system but also gives the true entropy for systems in string theory like the BIonic systems.

## 4 Summary and discussion

In this paper, we have obtained the entropy of an accelerating BIon which was constructed from two accelerating branes and a wormhole. We have observed that this entropy includes some terms similar to Tsallis entropy and could be written as $S = \gamma A^\beta$, where A is the area of BIon [1]. We have shown that by passing time and increasing the acceleration of BIon, the separation distance between branes and the length of wormhole grows. Consequently, the area of BIon increases and the entropy grows. This result is consistent with the second principle of thermodynamics, as well as with thermodynamical extensivity of the entropy.

**Acknowledgements** The work of Somayyeh Shoorvazi has been supported financially by the Research Institute for Astronomy and Astrophysics of Maragha (RIAAM),Iran under the Research Project NO.1/5237-91.